\documentclass{article}
\usepackage{icrctc07}
\title{Statistical Methods for Investigating the Cosmic Ray Energy Spectrum}
\shorttitle{The Cosmic Ray Energy Spectrum}
\authors{ J.~D.~Hague$^{1}$, B.~R.~Becker$^{1}$, M.~S.~Gold$^{1}$, J.A.J.~Matthews$^{1}$, J.~Urb\'{a}\v{r}$^{1}$. }
\shortauthors{J.~D.~Hague and et al}
\afiliations{$^1$University of New Mexico, Albuquerque, New Mexico USA }
\email{jhague@unm.edu}

\abstract{Two separate statistical tests are described and developed in order to test un-binned data sets
for adherence to the power-law form.
The first test employs the TP-statistic, a function defined
to deviate from zero when the sample deviates from the power-law form, regardless of the
value of the power index.
The second test employs a likelihood ratio test to reject a power-law background in favor 
of a model signal distribution with a cut-off.
}

\begin{document}
\maketitle
\newcommand{\pwlw}{{power-law }}
\newcommand{\tg}{{\gamma}}
\newcommand{\tgH}{{\hat{\gamma}}}
\newcommand{\tgHub}{{\hat{\gamma}}_{\text{\tiny ub}}}
\newcommand{\tgHlb}{{\hat{\gamma}}_{\text{\tiny lb}}}
\newcommand{\tgub}{{\gamma}_{\text{\tiny ub}}}
\newcommand{\otp}{{\overline{TP}}}

\section{Introduction and Formalism}
The question of whether the cosmic ray energy spectrum exhibits a cut-off at the very highest energies 
is of central interest to the cosmic ray (CR) physics\cite{refG,refZK}. 
The flux of CR's at these energies is very small - about $3 /$km$^{2}$ steradian century - 
and, therefore, statistical analysis techniques which clearly quantify ones knowledge of flux suppression are useful.
In this note we apply the statistics first developed for binned CR data sets in \cite{refHague} to an un-binned 
analysis. We also introduce a new test based on a likelihood ratio test and show that both statistics can 
quantify our knowledge of a flux suppression. 

We first establish the mathematical foundations of the analysis. 
The CR flux follows a power-law for over 10 orders of magnitude. 
The fundamental probability distribution function (p.d.f.) governing the power-law assumption 
(normalized such that $\langle \rangle_{X} \equiv \int_{x_{min}}^{\infty} f_{{\scriptscriptstyle X}}(x; x_{min}, \tg)dx=1$) is 
\begin{equation}\label{equpwlpdf}
f_{{\scriptscriptstyle X}}(x; x_{min}, \tg) = A\,x^{-\tg},  
\end{equation}
where $A=(\tg-1)x^{\tg-1}_{min}$ and the parameter $\tg$ is referred to as the {\it spectral index}.

The $n^{th}$ raw moment of this distribution diverges\cite{refNewm} for $n \geq 2$ with $\tg \leq 3$. 
Alternatively, the expected value of $\ln(x/x_{min})$ is better behaved and offers a crucial result of this analysis.
Analytically we find,
\begin{equation}\label{equnu}
\nu_{n}        \equiv \left\langle \ln^n \left( \frac{x}{x_{min}} \right) \right\rangle_{X} = \frac{n!}{(\tg-1)^{n}}. \,
\end{equation}
For a given sample we use,
\begin{equation} \label{equnuhat}
{\hat \nu}_{n}(X_{(j)}) \equiv \frac{1}{N-(j-1)} \sum_{i=j}^N \ln^{n} \left( \frac{X_{(i)}}{X_{(j)}} \right). \,  
\end{equation}
In eq.\ref{equnuhat} we denote the sorted (from least to greatest) 
data set as $\left\{X_{(1)},X_{(2)},\ldots,X_{(N)}\right\}$.
To apply these statistics to an un-binned data set we calculate ${\hat \nu}_{n}(X_{(j)})$ for each 
minimum $X_{(j)}$.   

We also study a toy p.d.f. which is designed to mimic a power-law up to a certain 
energy but then exhibit a sharp ``Fermi-Dirac like'' cut-off above that energy\cite{refHague}. 
We follow the parameterization used in \cite{refPAOicrc318},
\begin{equation}\label{equFDpdf}
f_{{\scriptscriptstyle FD}}(x; x_{c}, w_{c}, \tg) =  
              \frac{B\,x^{-\tg}}{1 + \exp \left( \frac{\log x - \log x_{c}}{ w_{c}} \right) },
\end{equation}
where $B$ is chosen such that 
$f_{{\scriptscriptstyle FD}}$ is normalized over the interval $[x_{min},\infty)$, i.e. 
$\langle \rangle_{{\scriptscriptstyle FD}} = 1$.

\section{Binned vs Un-binned Spectral-Index Estimators}
Under the power-law assumption, we can 
take the log of both sides of eq.\ref{equpwlpdf} to yield 
$\log f_{{\scriptscriptstyle X}} = \log((\tg-1)/x_{min}) - \tg \log(x/x_{min})$. 
The slope, $\tgHlb$, of the line which results 
in the minimum $\chi^{2}$ fit to the logarithmically binned (``LB'') histogram of a particular data. 
The un-binned maximum likelihood (``ub'') estimate of the spectral index can be found analytically\cite{refNewm}: 
\begin{equation}\label{equgub}
\tgHub(X_{(j)}) = 1 + 1 / {\hat \nu}_{1}(X_{(j)}).
\end{equation}

This estimator is within 1\% of the true $\tg$ for
$N \gtrsim 100$ and it is asymptotically unbiased.  
The variance of this estimator is within 1\% of the Cramer-Rao 
lower bound, given by $\sigma_{\tgH} \geq (\tg-1)/\sqrt{N}$, for\cite{refHowell} $N \gtrsim 100$. 
As derived in \cite{refHagueUB}, we write the asymptotic p.d.f. of $\tgHub$ as $f_{{\scriptscriptstyle ub}}(\tgub; N, \tg)$. 

\begin{figure}
\begin{center}
\includegraphics [width=65mm, height=35mm]{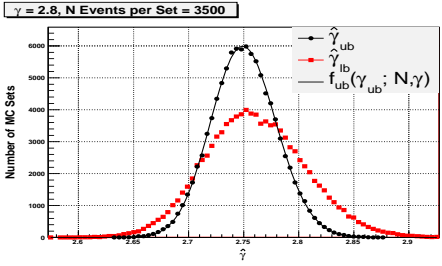} 
\end{center}
\caption{Estimates of the {\color{red}log-binned} ($\tgHlb$) and the {\color{black}un-binned} ($\tgHub$).
for $10^{5}$ Monte-Carlo trials. For each trail we draw $N=3500$ events from a power-law with $\tg=2.75$.
}\label{figSimGamma}
\end{figure}

To illustrate the benefits of using un-binned estimators $10^{5}$ Monte-Carlo trials were conducted.
For each trail we draw $N=3500$ events from a power-law with $\tg=2.75$ ($x_{min}=1$) and calculate 
$\tgHlb$ and $\tgHub$. 
These numbers are chosen to be approximately consistent with the flux reported\cite{refPAOicrc2005} 
by the Auger Collaboration at ICRC 2005, as studied in \cite{refHague}. 
In Figure \ref{figSimGamma} we plot histograms of these estimators and we note that 
the analytic prediction ($f_{{\scriptscriptstyle ub}}$ is not a ``fit'') represents a good approximation for the 
distribution of $\tgHlb$. 
The mean (over the trials) of $\tgHlb$ is $2.76$ with deviation $0.045$ while 
the corresponding values for $\tgHub$ are $2.75$ and $0.030$, verifying that $\tgHub$ has smaller error 
and less bias\cite{refGolds} than $\tgHlb$. 
Since we use 

\section{TP-statistic}
We define the TP-statistic to be,
\begin{eqnarray}
\tau        &=& \nu^{2}_{1} - \nu_{2}/2 = 0            \label{equtau} \\ 
{\hat \tau}(X_{(j)}) &=& {\hat \nu}^{2}_{1}(X_{(j)}) - \frac{1}{2}{\hat \nu}_{2}(X_{(j)}).    \label{equtauhat}
\end{eqnarray}
The utility of using this statistic comes from the fact\cite{refPisa1} that eq.\ref{equtau} is 
zero and thus, eq.\ref{equtauhat} will tend to zero as $N \rightarrow \infty$, regardless of the value of $\tg$. 

We may approximate the asymptotic joint distribution of ${\hat \nu}_{1}$ and ${\hat \nu}_{2}$ as 
a bivariate Gaussian $f_{{\scriptscriptstyle V_{1}V_{2}}}(\nu_{1}, \nu_{2})$ 
with known means, variances and correlation coefficient\cite{refHagueUB}.
Thus, for a given $N$ and $\tg$, we calculate the p.d.f. of $\tau$ to be,
\begin{equation}\label{equpdftau}
f_{{\scriptscriptstyle TP}}(\tau; N, \tg) = \int^{\infty}_{-\infty} 
                                            f_{{\scriptscriptstyle V_{1}V_{2}}}(t, 2(t^{2}-\tau))dt.
\end{equation}
The analytic ``location'' $\langle \tau \rangle_{{\scriptscriptstyle TP}}$ and ``shape'' 
$\langle \sigma_{\tau} \rangle_{{\scriptscriptstyle TP}} = 
\sqrt{ \langle\tau^{2}\rangle_{{\scriptscriptstyle TP}}- \langle \tau \rangle^{2}_{{\scriptscriptstyle TP}} }$ 
parameters of this distribution are consistent with simulation generated values.
Since the numeric integration required to calculate these quantities can be carried out faster than 
the requisite simulations we use the former to estimate the expected mean and variance of the 
power-law sample TP-statistic.

We estimate the significance of the TP-statistic for a given sample as 
\begin{equation}\label{equsigniftau}
(\hat{\tau} - \langle \tau \rangle_{{\scriptscriptstyle TP}})/\langle \sigma_{\tau} \rangle_{{\scriptscriptstyle TP}}.
\end{equation}
A spectrum with flux suppression in the tail (like that in eq.\ref{equFDpdf}) 
will result in a {\it positive} significance\cite{refHague}. 
We note from \cite{refHagueUB} that $\langle \sigma_{\tau} \rangle_{{\scriptscriptstyle TP}} \sim N^{-1/2}(\tg-1)^{-2}$.  

\begin{figure*}[t]
\begin{center}
\begin{tabular}{l r}
\includegraphics [width=65mm, height=35mm]{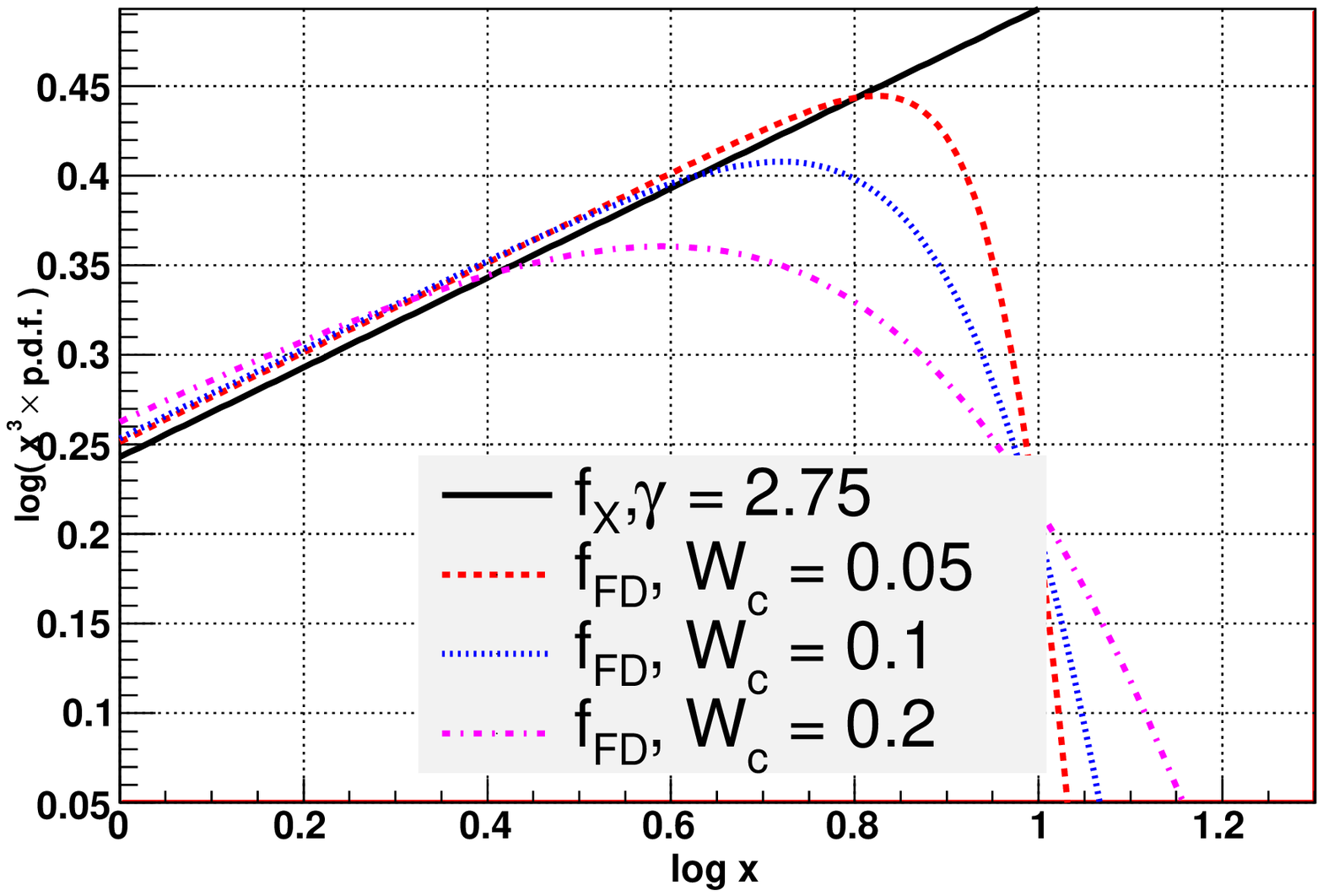} & 
\includegraphics [width=65mm, height=35mm]{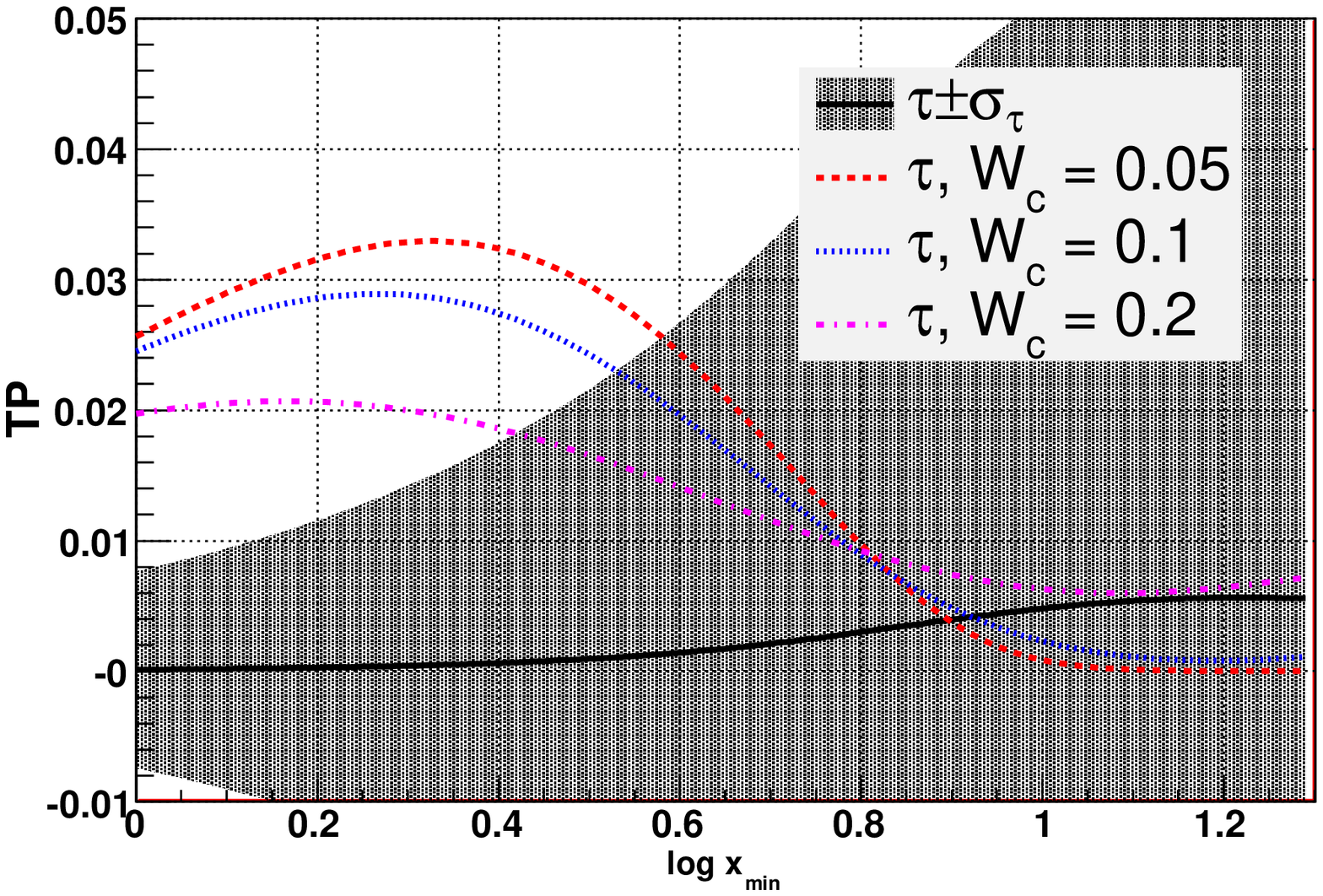}  \\
\includegraphics [width=65mm, height=35mm]{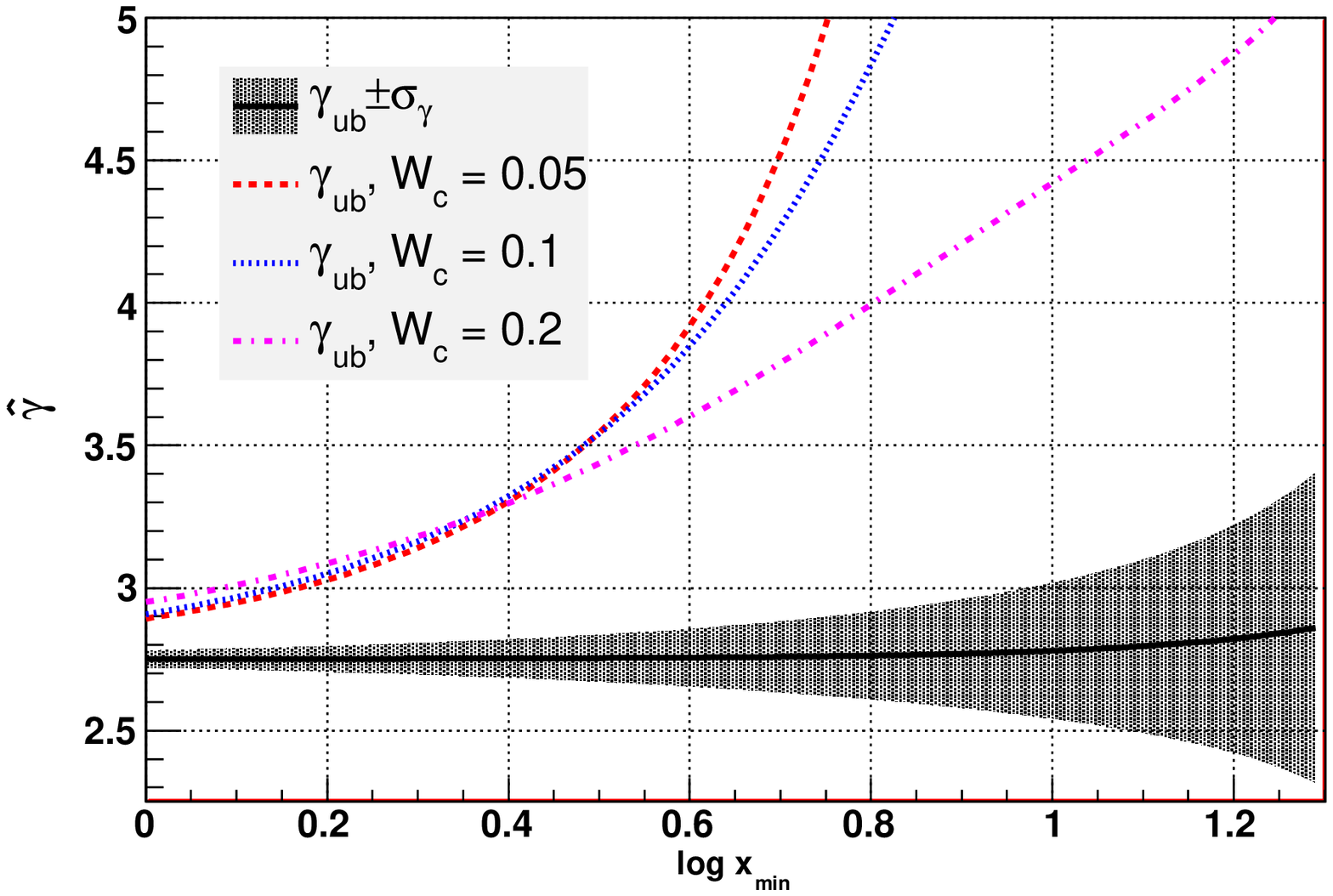} & 
\includegraphics [width=65mm, height=35mm]{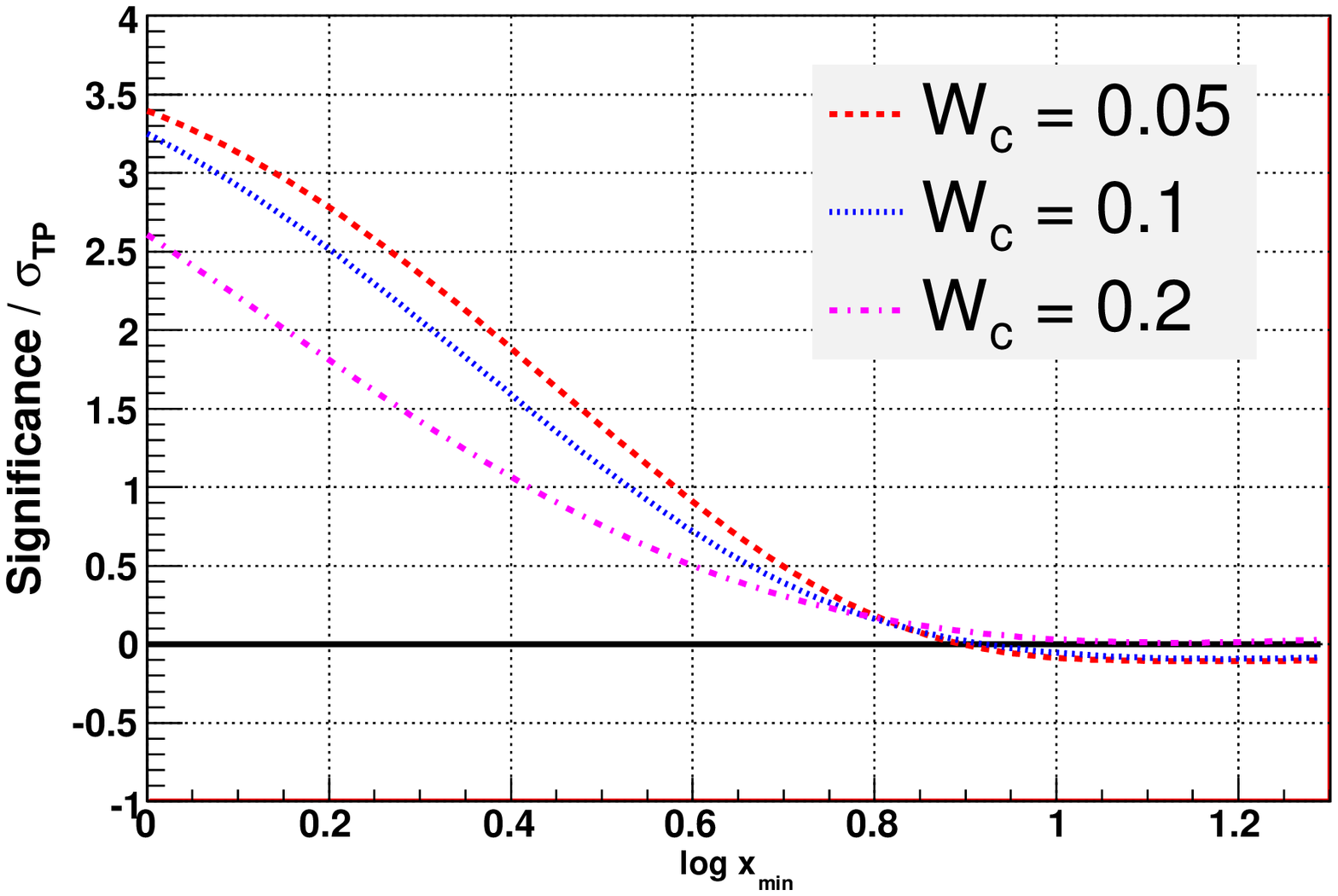}  
\end{tabular}
\end{center}
\caption{The TP-statistic is sensitive to flux suppression for these toy distributions, see text for explanation.}\label{figFermi}
\end{figure*}

In Figure \ref{figFermi} we illustrate the behavior of this statistic when applied to a distribution with suppression in the tail. 
Using eq. \ref{equFDpdf} we analytically calculate 
$\tgHub = 1 + \langle \ln(x/x_{min}) \rangle_{{\scriptscriptstyle FD}}$ 
(lower left) and 
$\tau = \langle \ln(x/x_{min}) \rangle^{2}_{{\scriptscriptstyle FD}} - 0.5\langle \ln^{2}(x/x_{min}) \rangle_{{\scriptscriptstyle FD}}$ 
(upper right) with $\tg=2.75$, $\log x_{c}=1.0$, for three choices of $w_{c}$ and as a function of $x_{min}$. 
We also calculate the expected value (and deviation) of these quantities when applied to a data set containing $3500$ events, 
drawn from a {\it pure power-law} with values greater than $1.0$ 
For each $x_{min}$ we estimate the number of events $N$ with value greater than $x_{min}$ as $3500x^{1-\tg}_{min}$.
The upper left panel shows the p.d.f.'s (on a log-log scale) normalized to unity on $[1.0, \infty)$.
The lower right contains the significance of the TP-statistic; for the lowest $x_{min}$ (i.e. $N=3500$) 
the model cut-off distributions can reject the power-law assumption at the $\sim 4\sigma$ confidence level.

\section{A Likelihood Ratio Test}
Here we introduce a likelihood ratio test designed to discriminate a model signal (power-law with a cut) from 
a background (pure power-law) hypothesis and to be weakly dependent on $\tg$.
We may write the natural log of the ratio of the signal likelihood 
$L_{{\scriptscriptstyle FD}} = \prod f_{{\scriptscriptstyle FD}}(x_{i})$ to that of the 
background $L_{{\scriptscriptstyle X}} = \prod f_{{\scriptscriptstyle X}}(x_{i})$ as,
\begin{eqnarray}\label{equlnRat}
&&R(\tg, \log x_{c}, w_{c}) = N \ln\left\{C(\tg, \log x_{c}, w_{c})\right\} \nonumber \\ 
&&- \sum_{i=1}^{N} \ln \left\{1+ \exp\left( \frac{\log x - \log x_{c}}{ w_{c}}\right) \right\}. 
\end{eqnarray}
We note that $C = B/A$ (see eqs. \ref{equFDpdf} and \ref{equpwlpdf}) contains the only dependence 
on $\tg$ and is independent of the data points 
under study, i.e. $R$ contains no term involving $\log x^{-\tg}_{i}$. 
Indeed, for any given $\log x_{c}$ and $w_{c}$, the quantity $\ln C$ is linearly dependent on $\tg$ with slope $\sim0.125$. 
In this sense the ratio test is weakly dependent on $\tg$. 
However, in order to evaluate the efficiency of this test to reject a particular power-law background in favor of the 
cut-off signal we must choose $\tg$ a priori. 

\begin{figure*}[t]
\begin{center}
\begin{tabular}{l c r}
\includegraphics [width=43.3mm, height=35mm]{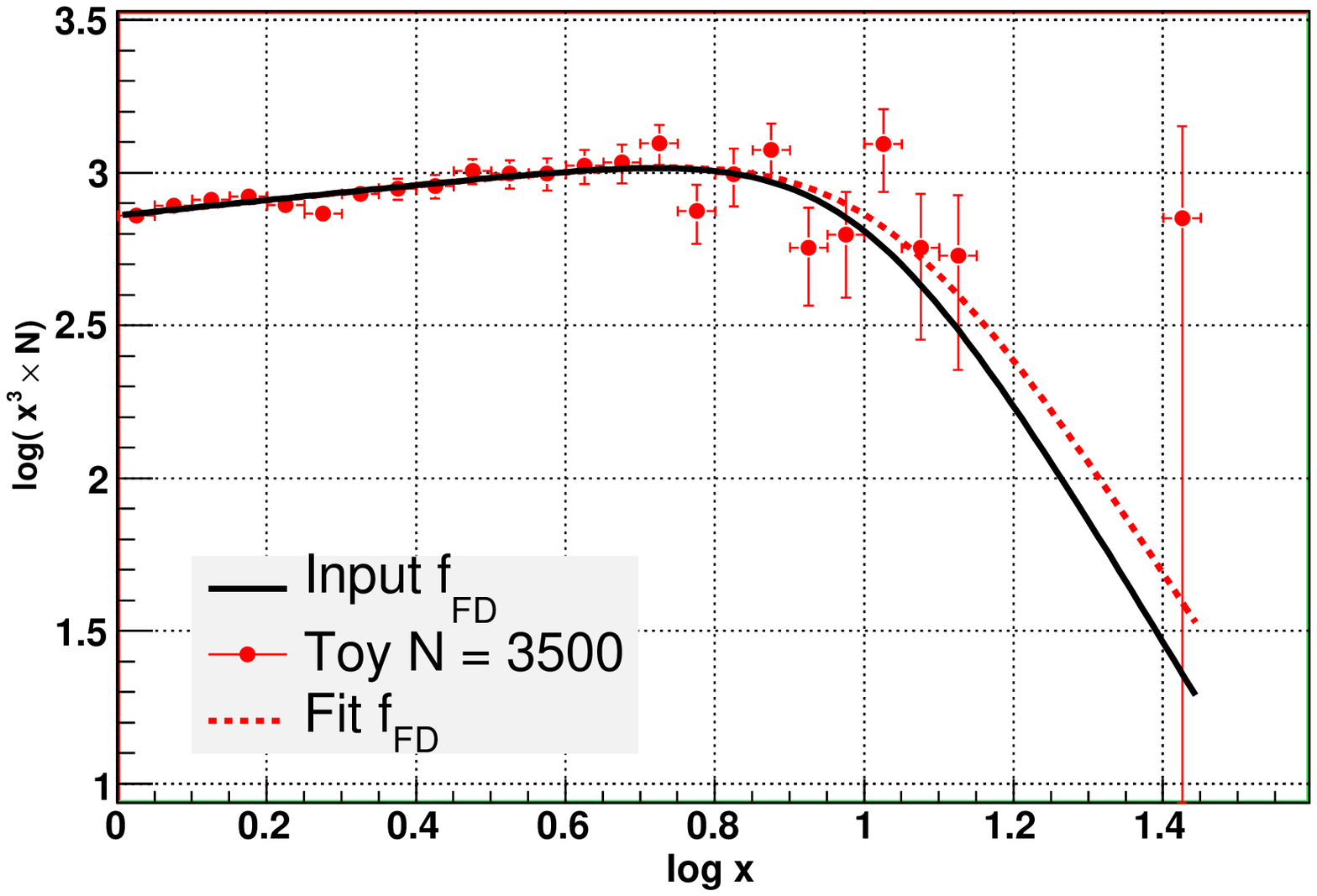} & 
\includegraphics [width=43.3mm, height=35mm]{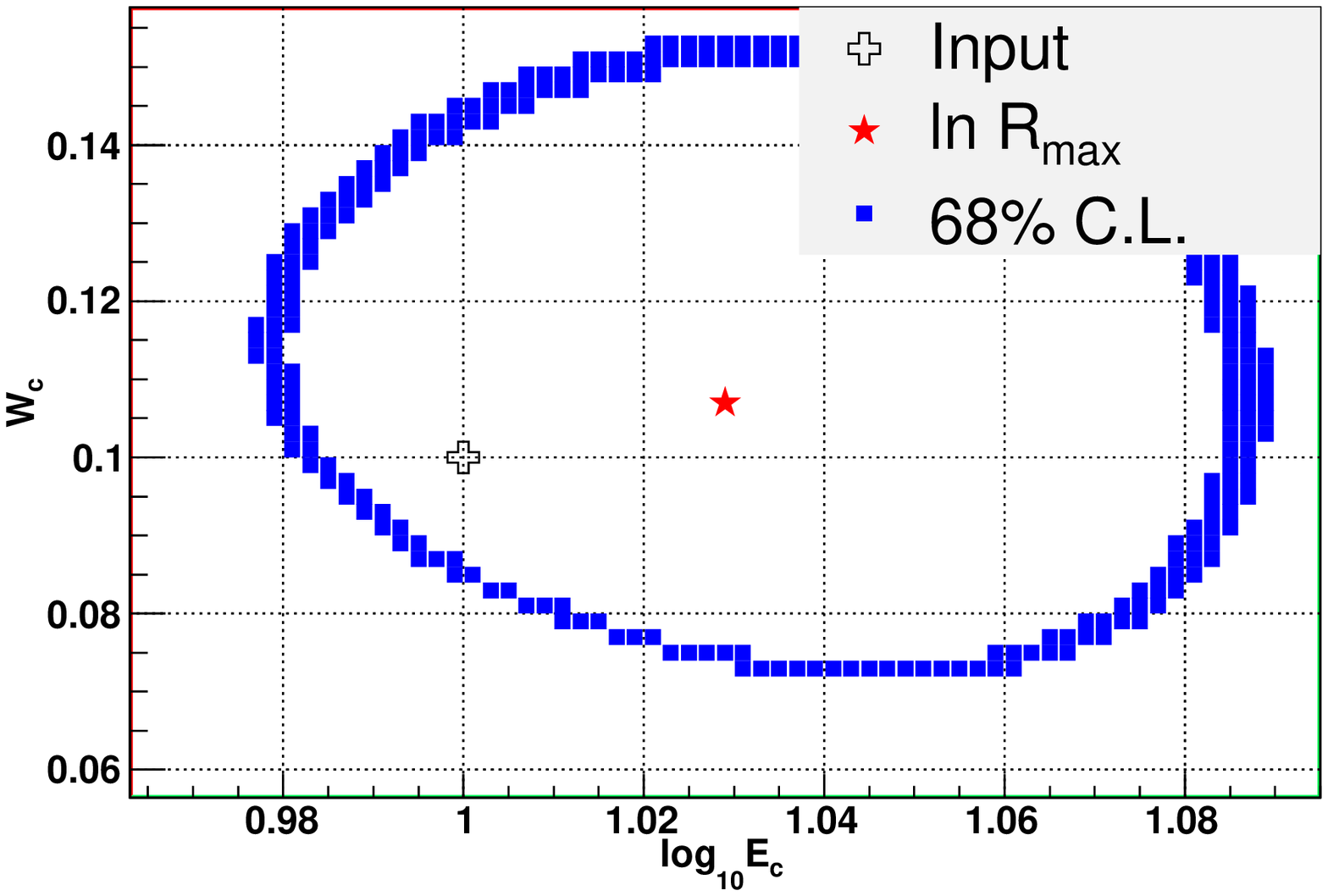} & 
\includegraphics [width=43.3mm, height=35mm]{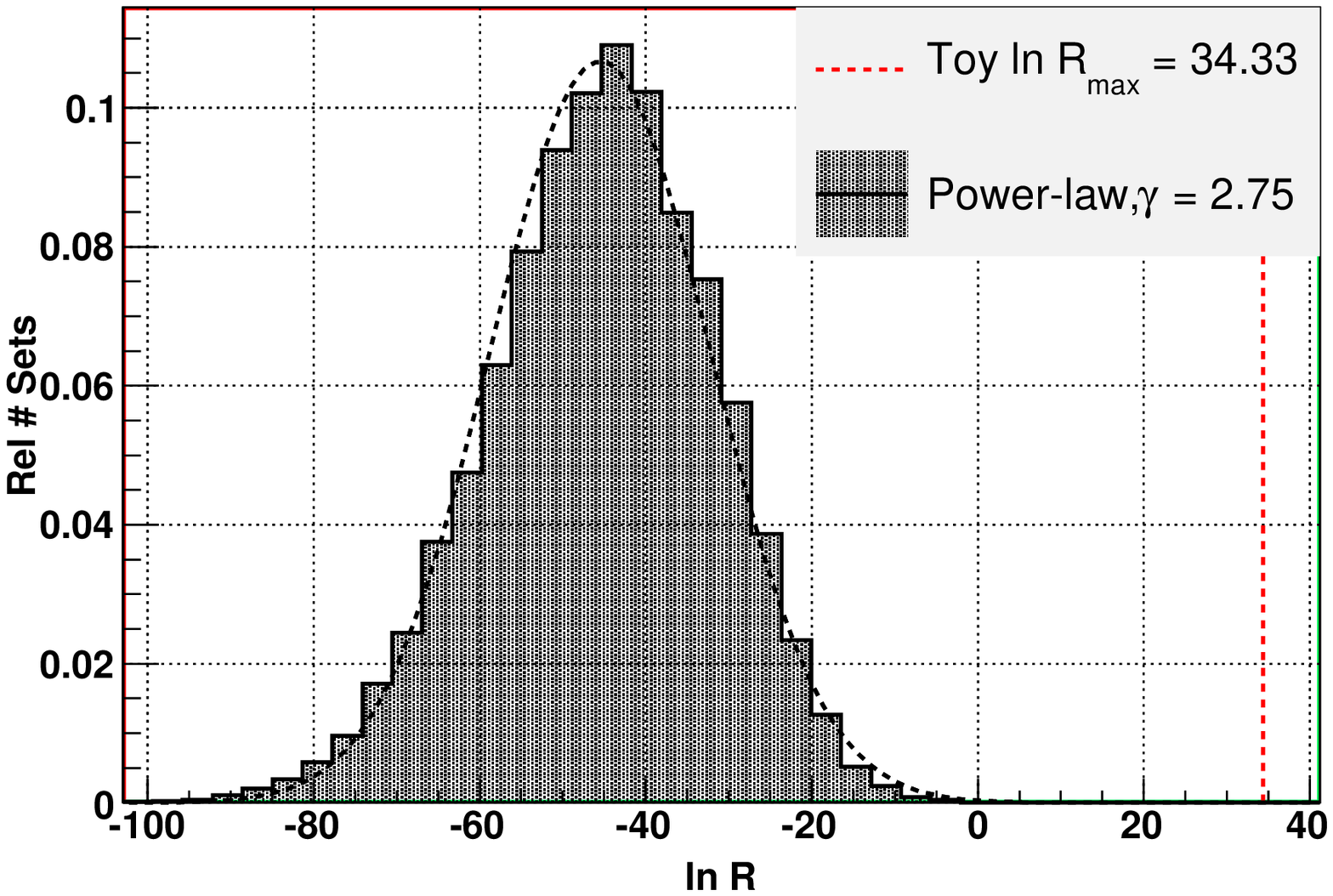} 
\end{tabular}
\end{center}
\caption{The ratio test is sensitive to flux suppression for this MC set, see text for explanation.}\label{figLNR}
\end{figure*}

To illustrate how this test could be applied to a CR data set we generate $3500$ ``toy'' events 
from $f_{{\scriptscriptstyle FD}}$ with input parameters $\tg=2.75$, $\log x_{c}=1$ and $w_{c}=0.1$ 
(see Figure \ref{figLNR}). 
With the a priori choice of $\tg=2.75$, we then calculate $R(2.75, \log x_{c}, w_{c})$ 
by scanning over the ranges $0.03 \leq w_{c} \leq 0.17$ and 
$0.93 \leq \log x_{c} \leq 1.07$. The maximum $\ln R_{max}=81.83$ gives us the fit parameter estimates 
$\log \hat{x}_{c} = 0.97 \pm 0.04$ and $\hat{w}_{c}=0.10 \pm 0.03$, where the $68\%$ confidence interval is 
approximated by the contour $\ln R_{max} - \ln R(2.75, \log x_{c}, w_{c}) = 2.30/2$. 

By simulating $N_{bg}=10^{4}$ sets of $3500$ background events drawn from a {\it pure power law} (with $\tg=2.75$) 
and performing the same parameter scan over $\log x_{c}$ and $w_{c}$, we can estimate the efficiency $\beta$ of 
this test to reject the power-law in favor the toy cut-off model, i.e. 
$\beta \sim N_{\ln R \geq \ln R_{max}}/N_{bg}$. 
From the right panel of Figure \ref{figLNR} we note that none of the $10^{4}$ background sets have $\ln R \geq \ln R_{max}$; 
we can reject the power-law in favor of the model cut-off at the $\sim4\sigma$ confidence level.

When applying this test to a real CR data set $\tg$ is not known a priori and one would want to estimate it. 
Studies of the ratio test with this extra degree of freedom are currently underway.

\section{Conclusions}
We began this note by verifying that the log-binned spectral index estimator has more bias and a larger 
error than the un-binned (maximum likelihood) estimator. 
We then detailed two un-binned statistical tests sensitive to flux suppression. 
We show that both tests 
show high sensitivity for rejecting the power-law hypothesis in favor of a toy flux suppression model 
and depend only weakly on the true spectral index. 
Applying these tests to $3500$ events drawn from a toy cut-off distribution (see eq. \ref{equFDpdf}) we 
can reject the power-law model in favor of the cut-off model at a confidence level $\sim 4$ standard deviations.

\bibliography{libros}
\bibliographystyle{plain}
\end{document}